\documentclass[prd,twocolumn,showpacs,amsmath,amssymb]{revtex4}
\usepackage{graphicx}

\newcommand{\jpsi}{J/\psi}
\newcommand{\psip}{\psi^{\prime}}
\newcommand{\psipto}{\psi^{\prime}\to}
\newcommand{\jpsito}{J/\psi\to}
\newcommand{\chic}[1]{\chi_{c#1}}
\newcommand{\chicto}[1]{\chi_{c#1}\to}
\newcommand{\etac}{\eta_{c}}
\newcommand{\etacp}{\eta_{c}^{\prime}}
\newcommand{\etacpto}{\eta_{c}^{\prime}\to}
\newcommand{\ee}{e^{+}e^{-}}

\newcommand{\piz}{\pi^{0}}

\newcommand{\pipi}{\pi^{+}\pi^{-}}
\newcommand{\kk}{K^{+}K^{-}}

\newcommand{\gam}{\gamma}
\newcommand{\gev}{~\mathrm{GeV}/c^{2}}
\newcommand{\mev}{~\mathrm{MeV}/c^{2}}

\newcommand{\GeV}{~\mathrm{GeV}}
\newcommand{\MeV}{~\mathrm{MeV}}

\def\Journal#1#2#3#4{{#1} {\bf #2}, #3 (#4)}

\def\PRL{Phys. Rev. Lett.}
\def\PRD{Phys. Rev. D}

\begin{document}

\title{\boldmath Search for $\etacp$ decays into vector meson pairs}
\author{
M.~Ablikim$^{1}$, M.~N.~Achasov$^{5}$, D.~Alberto$^{40}$,
F.~F.~An$^{1}$, Q.~An$^{38}$, Z.~H.~An$^{1}$, J.~Z.~Bai$^{1}$,
R.~Baldini$^{19}$, Y.~Ban$^{25}$, J.~Becker$^{2}$
% ...
\\
\vspace{0.2cm}
(BESIII Collaboration)\\
\vspace{0.2cm} {\it
$^{1}$ Institute of High Energy Physics, Beijing 100049, P. R. China\\
$^{2}$ Bochum Ruhr-University, 44780 Bochum, Germany\\
% ...
$^{5}$ G.I. Budker Institute of Nuclear Physics SB RAS (BINP), Novosibirsk 630090, Russia\\
% ...
$^{19}$ Laboratori Nazionali di Frascati - INFN, 00044 Frascati, Italy\\
% ...
$^{25}$ Peking University, Beijing 100871, P. R. China\\
% ...
$^{38}$ University of Science and Technology of China, Hefei 230026, P. R. China\\
% ...
$^{40}$ University of Turin and INFN, Turin, Italy\\
% ...
}}
\vspace{0.4cm}

\date{\today}

\begin{abstract}

    The processes $\etacpto \rho^{0}\rho^{0}$, $K^{*0}\bar{K}^{*0}$, and
    $\phi\phi$ are searched for using a sample of $1.06 \times 10^8$
    $\psip$ events collected with the BESIII detector at the BEPCII
    collider. No signals are observed in any of the three final
    states. The upper limits on the decay branching fractions are
    determined to be
    $\mathcal{B}(\etacpto\rho^{0}\rho^{0})<3.1\times10^{-3}$,
    $\mathcal{B}(\etacpto K^{*0}\bar{K}^{*0})<5.4 \times10^{-3}$, and
    $\mathcal{B}(\etacpto\phi\phi)<2.0\times10^{-3}$ at the 90\%
    confidence level.  The upper limits are lower than the existing
    theoretical predictions.

\end{abstract}

\pacs{14.40.Pq, 12.38.Qk, 13.20.Gd, 13.25.Gv}

\maketitle

The radially ($n$=2) excited $S$-wave spin-singlet charmonium
state, $\etacp$, labeled $\eta_{c}(2S)$, was observed in $B^{\pm}
\to K^{\pm}\etacp$, $\etacp \to K^{0}_{S}K^{\pm}\pi^{\mp}$ by the
Belle Collaboration~\cite{etacp_belle} and was confirmed by the
CLEO and BaBar collaborations~\cite{etacp_babar_cleo}. In addition
to the $K\bar{K}\pi$ final state, $\etacpto 3(\pipi)$,
$\kk2(\pipi)$, $K_{S}^{0}K^{\pm}\pi^{\mp}\pipi$, and
$\pipi\kk\piz$ are also reported~\cite{etacp_ichep}.  The
production of $\etacp$ is also expected from the radiative
magnetic dipole~($M1$) transition of $\psip$. The decay $\psipto
\gam\etacp$, $\etacpto K^{0}_{S}K^+\pi^-+c.c.$ was observed at
BESIII~\cite{etacp_bes3} with a branching fraction
$\mathcal{B}(\psipto\gamma\etacp) = (4.7\pm 0.9\pm 3.0)\times
10^{-4}$, confirming the possibility to study $\etacp$ properties
in $\psip$ transitions. In this analysis, we search for the
$\etacp$ decaying into vector meson pairs.

The decay modes $\etacpto VV$, where $V$ stands for a light vector
meson, are supposed to be highly suppressed by the helicity
selection rule~\cite{hsr_1}. But in Ref.~\cite{theo_VV}, a
higher production rate of $\etacpto VV$ is predicted, taking into
consideration significant contributions from intermediate charmed
meson loops, which provide a mechanism to evade
helicity selection rule~\cite{theo_VV_1}. The intermediate charmed meson loops can
also significantly suppress $\psipto VP$ (where $P$ stands for a
pseudoscalar meson) strong decay amplitudes~\cite{rhopi}, which
may help to explain the ``$\rho\pi$ puzzle'' in charmonium
decays~\cite{rhopi_review}. The measurement of
$\mathcal{B}(\etacpto VV)$ may help in understanding the role
played by charmed meson loops in $\etac\to VV$.

In this study, an $\ee$ annihilation data sample with
$(1.06\pm0.04)\times10^{8}$ $\psip$ events~\cite{n_psip} is
analyzed. Another data sample of $923~\mathrm{pb}^{-1}$ at
$\sqrt{s}=3.773\GeV$ is used to estimate non-$\psip$ background.
The data were collected with the BESIII detector which is
described in detail elsewhere~\cite{bes3_detector}. A charged-particle 
tracking system, main drift chamber, is immersed in a 1 T magnetic field.
A time-of-flight system and an
electromagnetic calorimeter (EMC) surrounding the tracking system
are used to identify charged particles and to measure neutral
particle energies, respectively. Located outside the EMC, a muon
chamber is used to detect muon tracks.

A Monte Carlo (MC) simulation is used to determine the mass
resolution and detection efficiency, as well as to study
backgrounds. The simulation of the BESIII detector is based on
\textsc{geant4}~\cite{geant}, where the interactions of particles
with the detector material are simulated. We use the program
\textsc{lundcrm}~\cite{lundc} to generate inclusive MC events for
the background study, where the branching fractions for known
decay channels are taken from the Particle Data Group
(PDG)~\cite{PDG2010}. For the signal channel $\psipto \gam\etacp$,
the photon is generated with the polar angle distribution
$1+\cos^{2}\theta$. To generate the correct decay angle
distributions, the $\etacpto VV$ decays are modeled with SVV
model~\cite{svv_model}, and $V$ decays are generated by the VSS
model~\cite{EvtGen}, which is used to describe decays of a vector
particle into two scalars.

We search for the $\etacp$ in three exclusive decay channels:
$\psipto \gam\rho^{0}\rho^{0}\to \gam2(\pipi)$, $\psipto \gam
K^{*0}\bar{K}^{*0}\to \gam\pipi\kk$, and $\psipto \gam\phi\phi\to
\gam2(\kk)$.  These final states, denoted as $\psipto \gam X$
hereafter, contain one radiative photon and four charged tracks.
The charged tracks are required to pass within 1~cm of the $\ee$
annihilation interaction point transverse to the beam line
and within 10~cm of the interaction point along the beam axis. Each track should
have good quality in track fitting and satisfy
$|\cos\theta|<0.93$, where $\theta$ is the polar angle with
respect to the $e^{+}$ beam direction. Reconstructed events are
required to have four charged tracks and zero net charge.
Information from $dE/dx$ and time-of-flight is used for charged-particle
identification (PID), and $\chi^{2}_{PID}~(i)$ is calculated for
each charged track, where $i$ is the corresponding
charged-particle hypothesis including pion, kaon, and proton.
For a specific decay channel, the total $\chi^{2}_{PID}$ is obtained
by summing $\chi^{2}_{PID}~(i)$ over the charged tracks. There is
a loop to match the charged tracks to the final state particles in the
decay channel, and the matching with the minimum $\chi^{2}_{PID}$ is
adopted. The decay channel for a reconstructed event is selected as
the one with the minimum $\chi^{2}_{PID}$ among possible decay channels.
Photons are reconstructed by clustering
EMC crystal energies with a minimum energy of 25$\MeV$.
The photon candidates are required to be detected in the active area
of the EMC ($|\cos\theta_{\gam}|< 0.8$ for the barrel and
$0.86<|\cos\theta_{\gam}|<0.92$ for the endcaps). Timing
requirements are used in the EMC to suppress electronic noise and
energy deposits unrelated to the event.

In order to reduce background from non-$VV$ production, the
invariant masses of the final decay particles are required to
satisfy $0.67\gev<M_{\pipi}<0.87\gev$,
$0.85\gev<M_{\pi^{\pm}K^{\mp}}<0.95\gev$, and
$1.01\gev<M_{\kk}<1.03\gev$, for $\rho^{0}$, $K^{*0}$ and $\phi$ candidates,
respectively, which are determined by fitting their mass
distributions in the $\chic{J}$ mass region.
Here the background level has been considered in
the choice of the selection criterion for each channel. The ratios
of signal over non-$V$ background are near 1 at the edges of the mass selection
region for $\rho^{0}$ and $K^{*0}$.

A kinematic fit is performed to improve the mass resolution and
reject backgrounds. The four-momenta of the charged tracks and the
photon candidate are constrained to the initial $\psip$
four-momentum~(4C fit). When there is more than one photon, the
photon with the minimum $\chi^{2}$ from the 4C fit,
$\chi^{2}_{4C}$, is taken as the radiative photon, and $\chi^{2}_{4C}$
is required to be less than 40.

Background from $\psipto \pipi\jpsi$ with $\jpsi$ decaying into a
lepton pair is removed by requiring the recoil mass~\cite{rec_mass} of any $\pipi$
pair to be below the $\jpsi$ mass~($m_{\pipi}^{\rm recoil} <
3.05\gev$). Events from $\psipto \eta\jpsi$, with $\eta\to
\pipi\piz~(\gam)$ and $\jpsi$ decays into lepton pairs, are also
removed by this requirement.

The background remaining can be separated into three categories:
events with no radiative photon ($\psipto
X$); events with an extra photon in the final state ($\psipto \piz X$,
$\piz\to \gam\gam$); and events with the same final state as the signal
($\psipto\gam X$), but where the photon comes from initial state
radiation or final state radiation ($FSR$).

\begin{figure}[htbp]
    \begin{center}
        \includegraphics[width=0.45\textwidth]{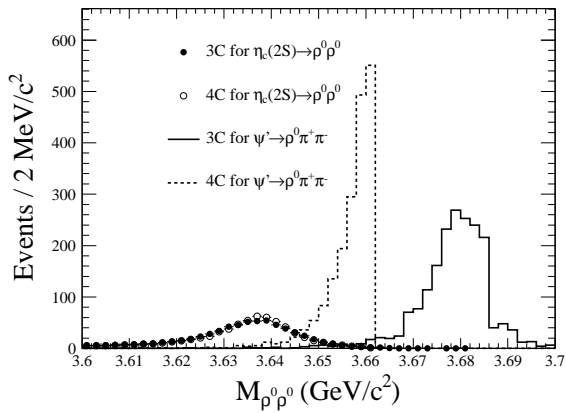}
        \caption{\label{mass_3C4C} Comparison between 3C and 4C kinematic fits
        (unnormalized). Shown in the plot are the signal with the 3C fit
        (filled circles), signal with the 4C fit (open circles), $\psipto X$
        background with the 3C fit (solid line), and $\psipto X$ background
        with the 4C fit (dashed line).}
    \end{center}
\end{figure}

The background from $\psipto X$ with no radiative photon comes
from events where the charged tracks plus a fake photon satisfy
the 4C kinematic fit. In the $X$ mass spectrum from a 4C kinematic
fit, this background contributes a peak close to the $\etacp$
mass, around $3.656\gev$, and decreases sharply at high mass due
to the $25\MeV$ requirement on the photon energy. If the measured
energy of the candidate photon is not used in the kinematic fit,
thus becoming a 3C fit, this background lies around the $\psip$
mass region~($3.66\gev\sim 3.70\gev$) in the mass spectrum, as the
photon energy from the fit tends to be close to zero energy (see
Fig.~\ref{mass_3C4C}). There is little change in the $\etacp$ mass
resolution due to one less constraint in the kinematic fit, but
the separation of the $\etacp$ signal from the background is much
improved. Therefore, the result from the 3C fit ($M_{X}^{3C}$) is
taken as the final mass spectrum.

\begin{figure*}[htbp]
    \begin{center}
        \includegraphics[width=0.450\textwidth]{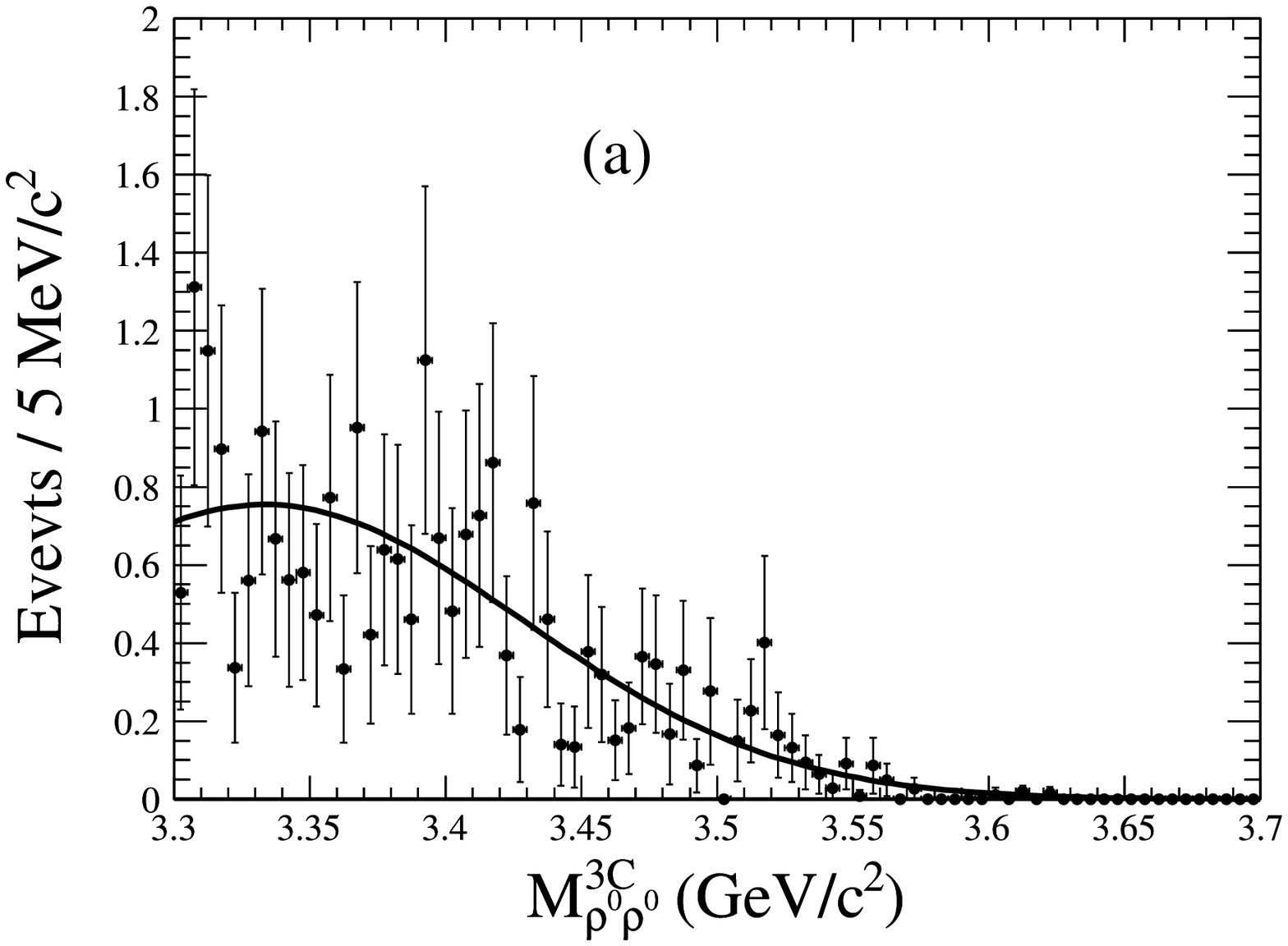}%
        \includegraphics[width=0.450\textwidth]{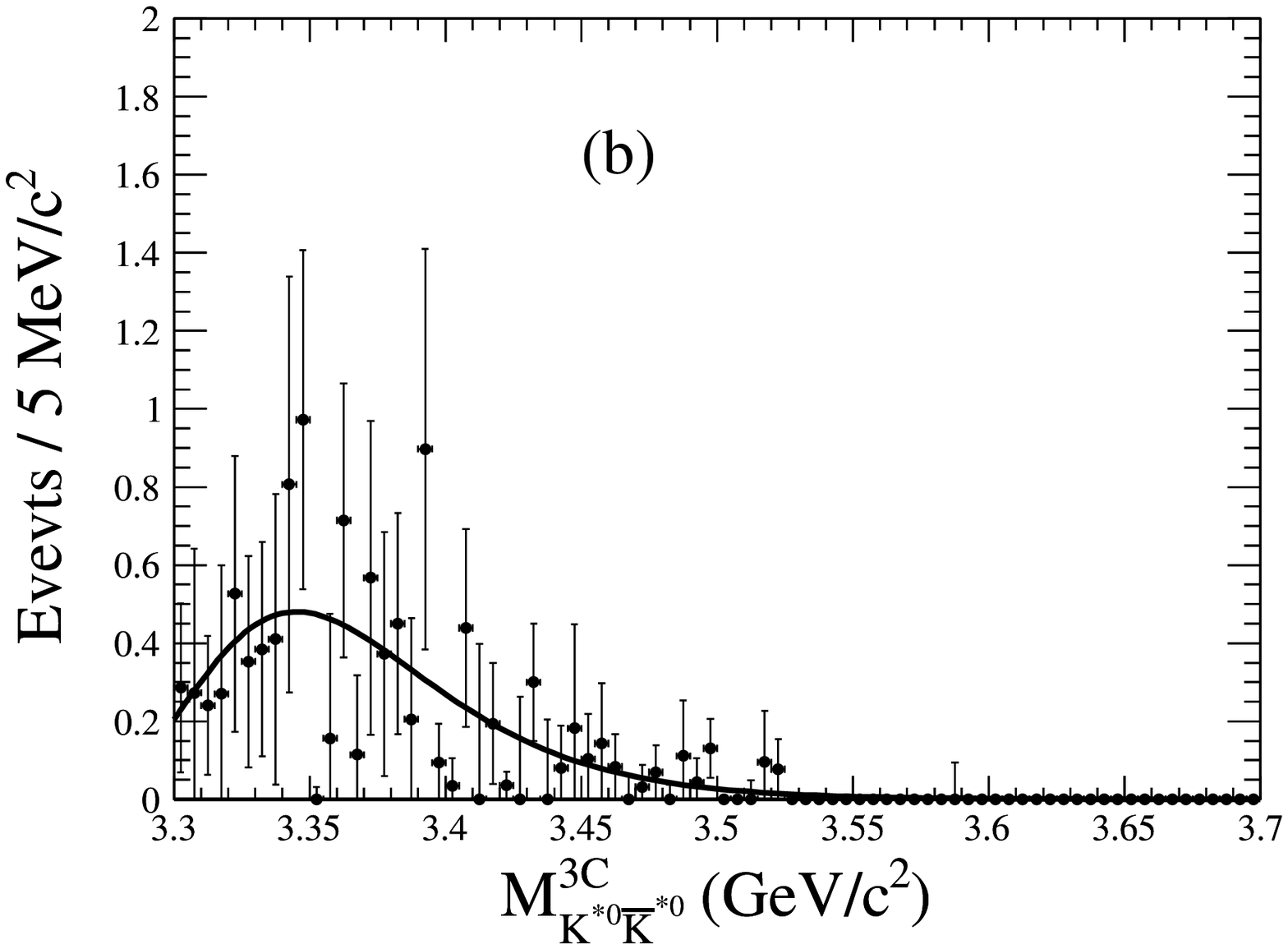}%
        \caption{\label{fit_pi0}The measured background from $\psipto\piz X$
        events (dots with error bars) for the modes: (a)
        $\gamma\rho^{0}\rho^{0}$ and (b) $\gamma K^{*0}\bar{K}^{*0}$.  The
        curves show the best fit with Novosibirsk functions.}
    \end{center}
\end{figure*}

The background from $\psipto\piz X$ is measured from data by
reconstructing the $\piz$ from its decay into two photons.
If there are more than two photons, the $\piz$ candidate is
selected as the one with the minimum $\chi^{2}$ from a 5C fit (4C
plus a $\piz$ mass constraint). $\chi^{2}_{5C}<30$ is required to
veto backgrounds. A MC sample of
$\psipto \piz X$ is used to determine the efficiency ratio between
events passing the $\psipto\gam X$ and $\psipto\piz X$ selections.
Finally, the efficiency ratio is used to scale the $\psipto\piz X$
sample selected from data to obtain the background contamination
from $\psipto\piz X$ as a function of the $X$ invariant mass.
This background, which is described with a
Novosibirsk function \cite{f_Novo} as shown in Fig.~\ref{fit_pi0},
contributes a smooth component in the $\chic{J}$ ($J=0$, 1, 2)
mass region ($3.35\gev\sim 3.60\gev$), and is almost negligible
above $3.60\gev$.

\begin{figure*}
    \begin{center}
        \includegraphics[angle=0, width=0.33\textwidth]{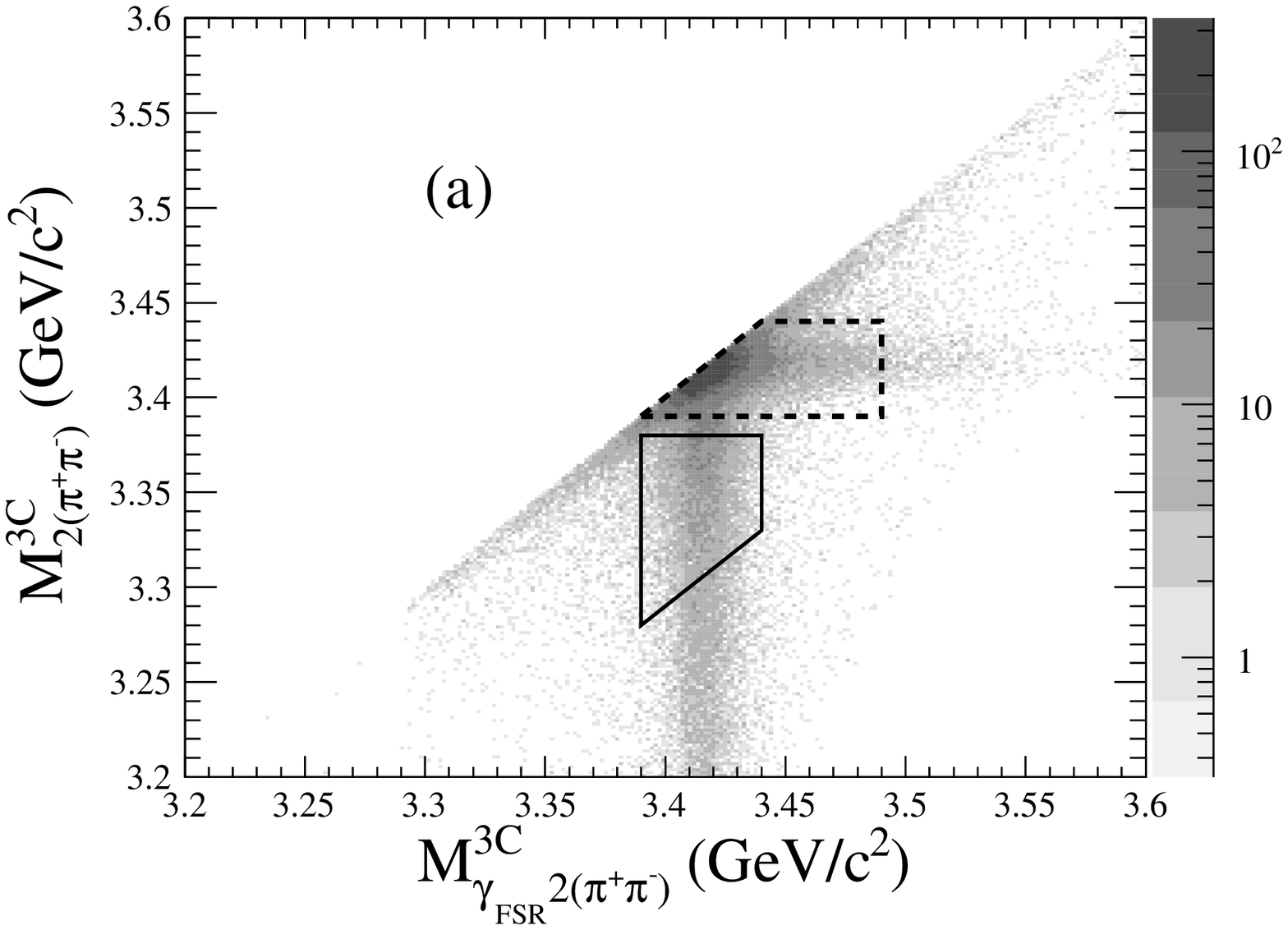}%
        \includegraphics[angle=0, width=0.33\textwidth]{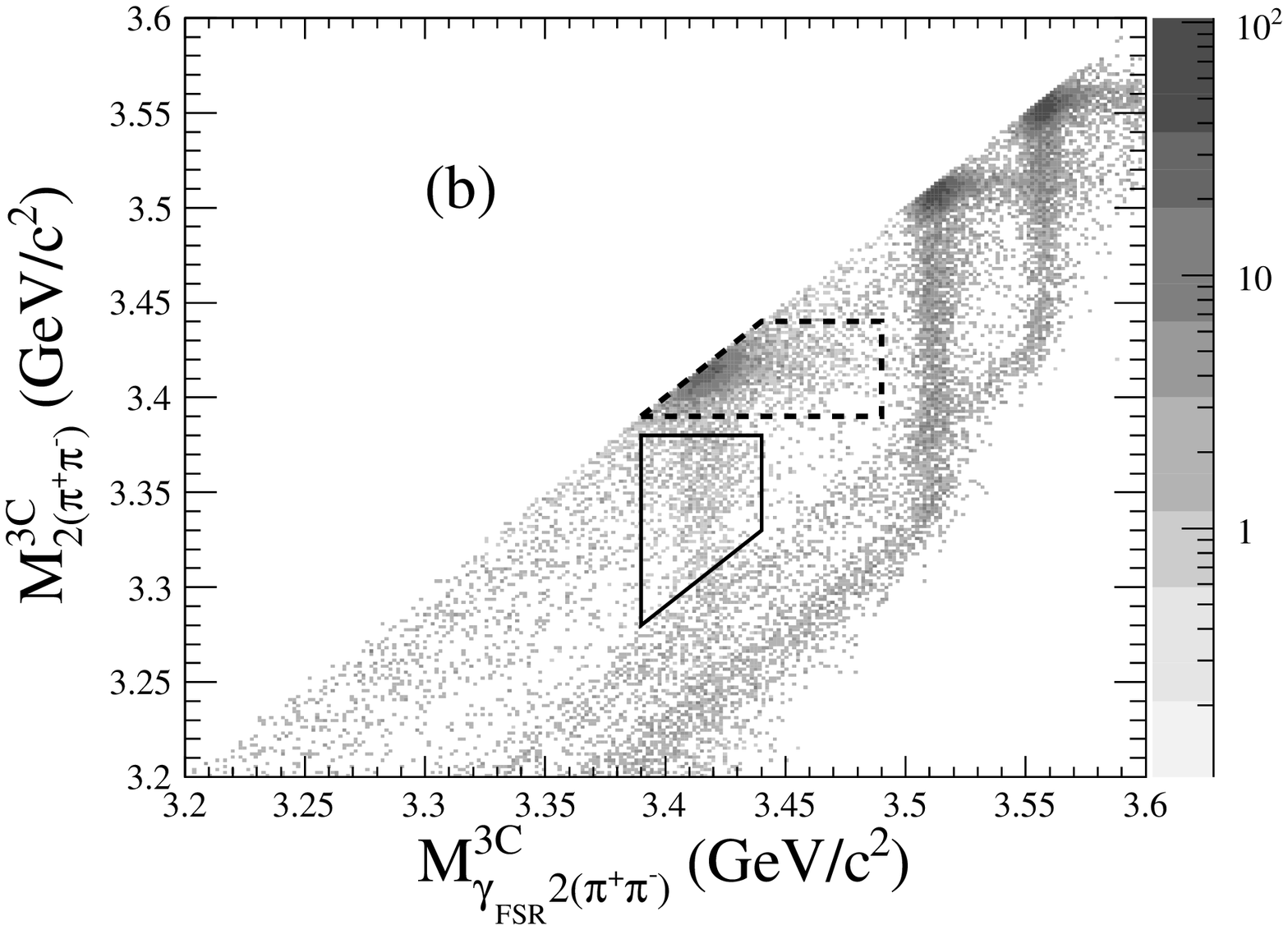}%
        \includegraphics[angle=0, width=0.33\textwidth]{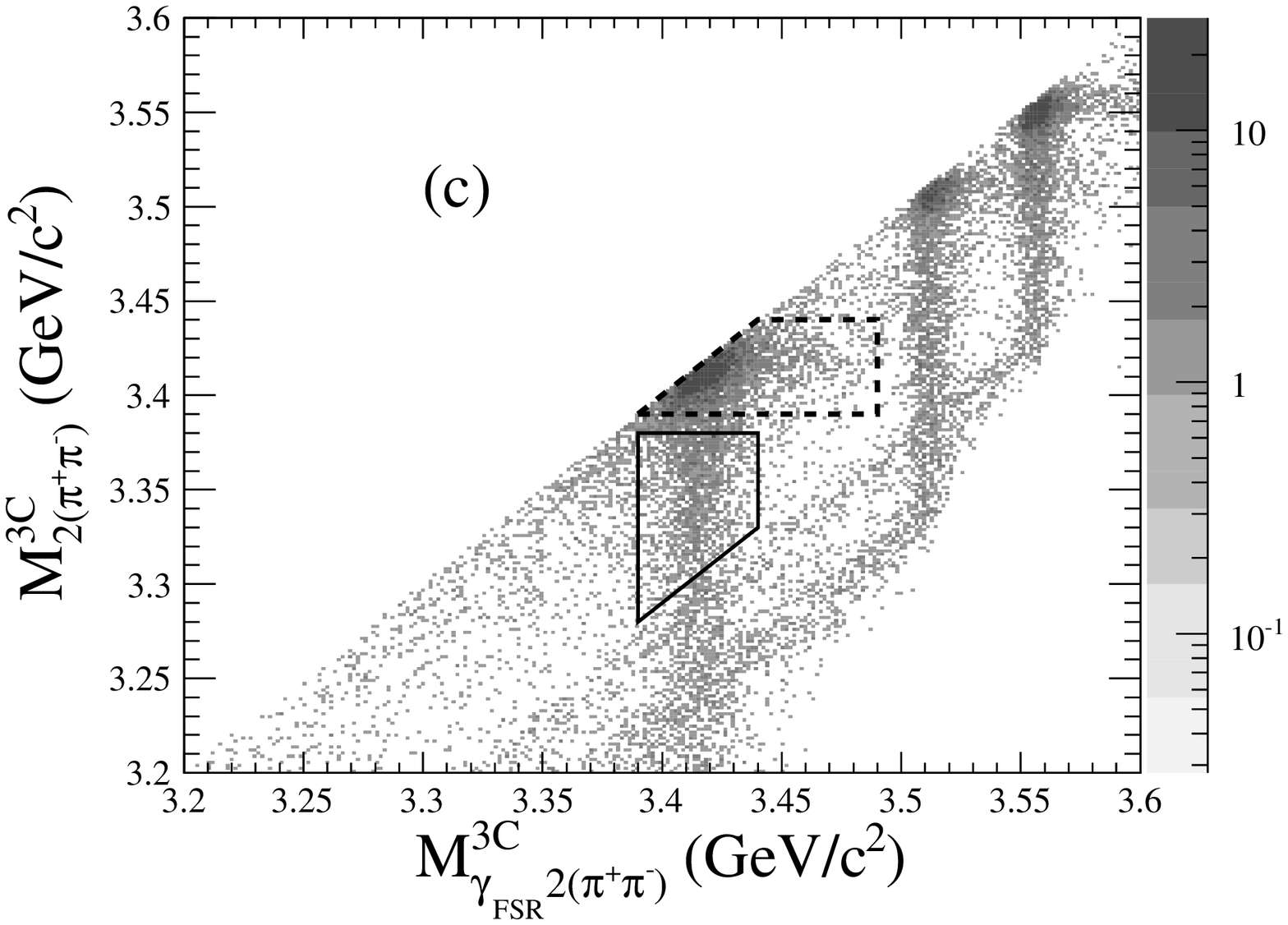}%
        \caption{\label{m2D}The two-dimensional plots of $M_{X}^{3C}$ versus
        $M_{\gam_{FSR}X}^{3C}$ for events passing the $\psipto\gam\gam_{FSR}X$
        selection with $X=2(\pipi)$. From left to right they are (a) MC
        simulated $\chic{0}$ signal, (b) inclusive MC, and (c) data. In each plot the
        dashed-line and the solid-line boxes contain events without and with a
        $FSR$ photon, respectively. MC simulations reproduce the shape well
        but not the amount of $FSR$ events.}
    \end{center}
\end{figure*}

The background shape from $\psipto(\gam_{FSR}) X$ is obtained from
MC simulation, where the $FSR$ photon is simulated with
PHOTOS~\cite{photos}. The fraction of events with $FSR$ is defined
as $R_{FSR}= \frac{N_{\gam_{FSR} X}}{N_{X}}$, where $N_{\gam_{FSR} X}$
($N_{X}$) is the number of events containing an (no) $FSR$ photon
that survive selection. This fraction is obtained
from measuring the $FSR$ contribution in $\psipto \gam\chic{0}$,
$\chicto{0}(\gam_{FSR}) X$. The event selection of this $FSR$
sample is very similar to that of the signal mode, except that
the reconstructed final state contains two photons, where the softer
photon is regarded as the $FSR$ photon. The energy of the $FSR$ photon is
not used when performing the 3C kinematic fit for this sample. Events from
$\psipto\piz X$ are the main background for the $FSR$ sample and
are excluded by requiring the invariant mass of the two photons to be
outside of the $\piz$ signal region. Figure~\ref{m2D} shows the two-dimensional
distribution of $M_{X}^{3C}$ versus $M_{\gam_{FSR}X}^{3C}$. If we add the four-momenta of the
$FSR$ photon and $X$ to calculate the invariant mass for events with $M_{X}^{3C}$
below the $\chic{0}$ mass in the PDG ($M_{\chic{0}}^{PDG}$), $M_{\gam_{FSR}X}^{3C}$ peaks at
$M_{\chic{0}}^{PDG}$ indicating the photon is indeed from $FSR$. As a result,
events from $\chicto{0}X$ are in the dashed-line box in
Fig.~\ref{m2D}, while events from $\chicto{0}\gam_{FSR}X$ are in
the solid-line box in Fig.~\ref{m2D}. In this way, we can
obtain $R_{FSR}$ for MC simulation and data. The factor
$f_{FSR}$ is defined as the ratio of $R_{FSR}$ measured in data to
that determined in MC simulation. This $FSR$ measurement is performed
for two final states; $f_{FSR}=1.70\pm 0.10$ and $1.39\pm 0.08$
are determined for $X=2(\pipi)$ and $X=\pipi\kk$, respectively.
The errors are the statistical errors of the sample and the
uncertainties of the background estimation. These factors are used
to scale fractions of $FSR$ background events
[$\psipto(\gam_{FSR})X$] in the MC samples to estimate the
background in data.

Data taken at $\sqrt{s}=3.773\GeV$ are used to estimate
backgrounds from the continuum [$\ee\to\gam^{*}\to(\gam_{FSR}) X$]
and initial state radiation ($\ee\to\gam_{ISR} X$). MC simulation indicates that
$\psi''$ decays contribute negligible background in the modes
under study. Using the luminosity
normalization and energy dependence of the cross section, there
are $46\pm 3$ and $8\pm 2$ background events expected for
$V=\rho^{0}$ and $V=K^{*0}$, respectively. For $V=\phi$, no events
survive the selection.

\begin{figure}[htbp]
    \begin{center}
        \includegraphics[width=0.5\textwidth]{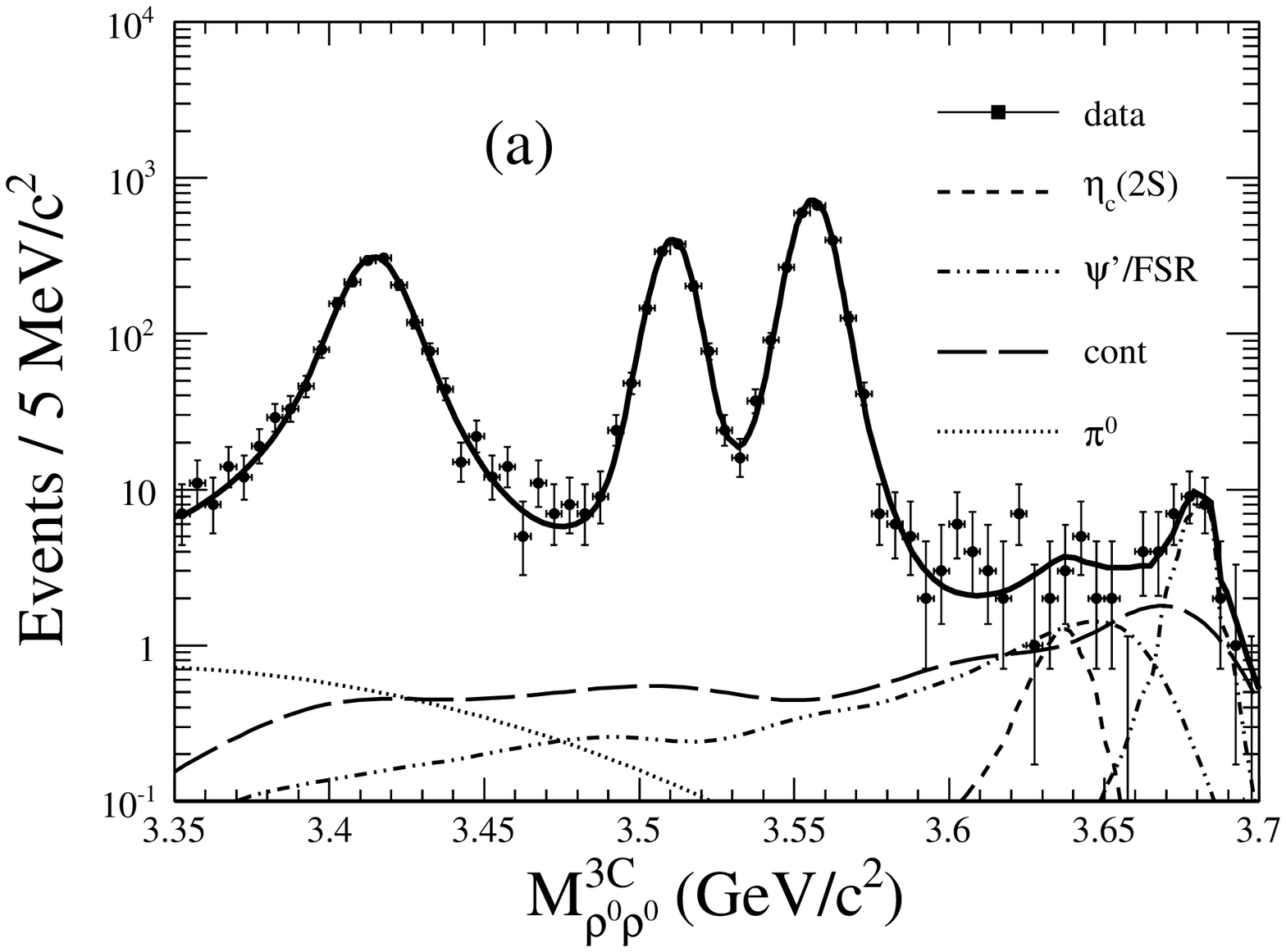}\\%
        \includegraphics[width=0.5\textwidth]{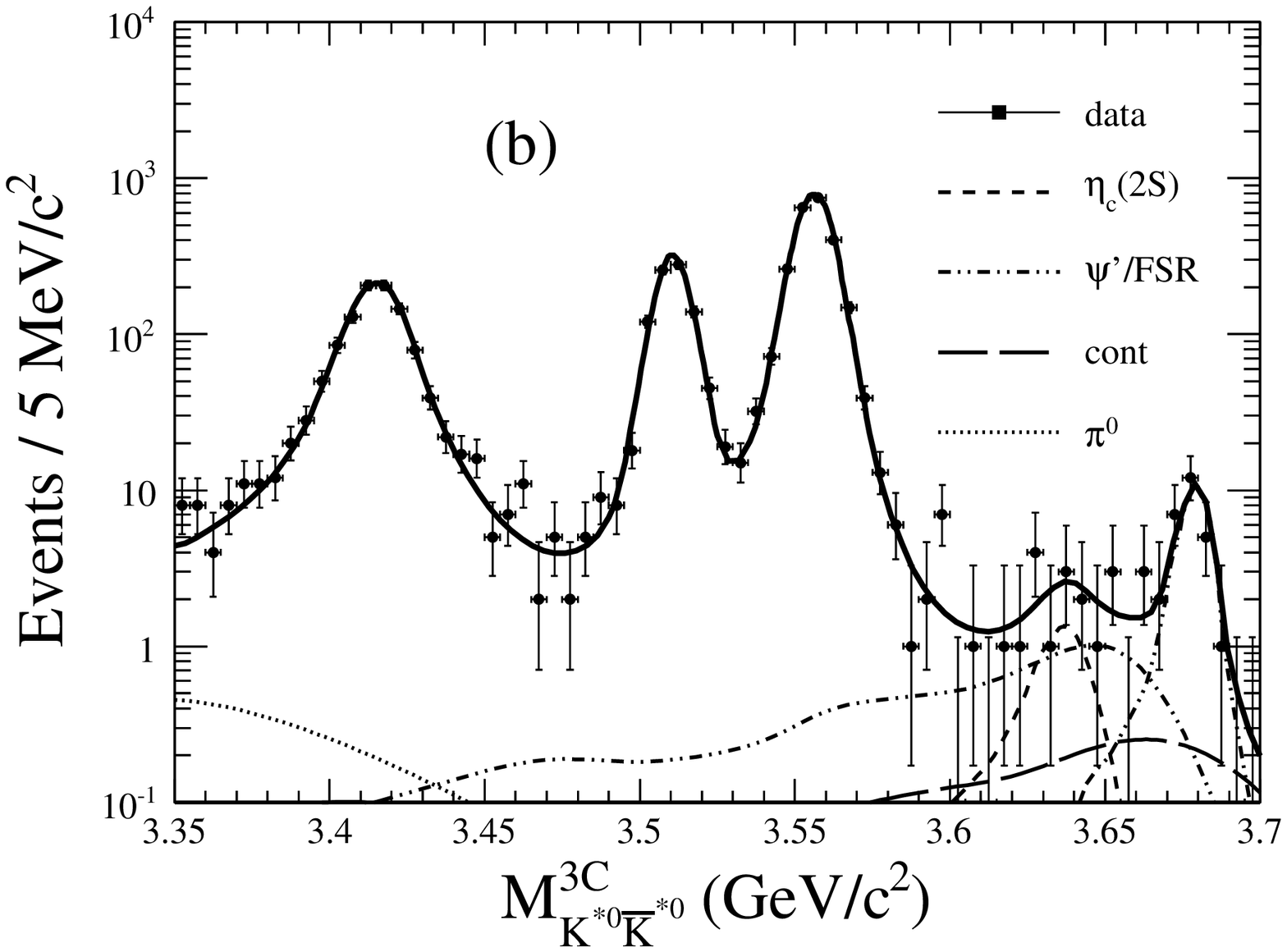}\\%
        \includegraphics[width=0.5\textwidth]{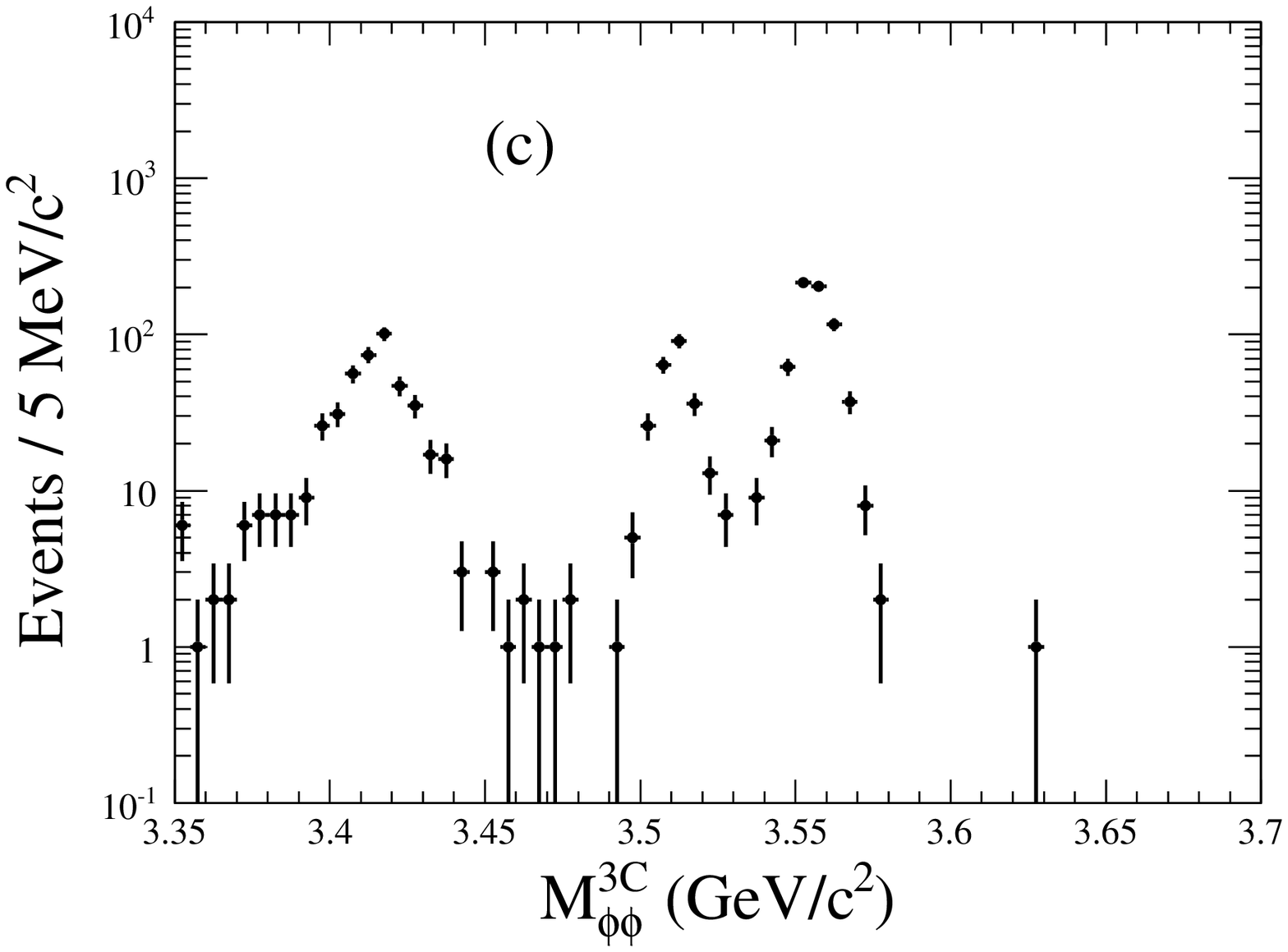}%
        \caption{\label{fit_VV}Invariant mass distributions of the vector
        meson pairs after a 3C kinematic fit for the modes (a)
        $\rho^{0}\rho^{0}$, (b) $K^{*0}\bar{K}^{*0}$, and (c) $\phi\phi$.  Dots
        with error bars are data, and the solid curves in (a) and (b) are from
        the best fit to the mass spectra. No fit is performed for (c) due to
        low statistics. In (a) and (b), the $\etacp$ signals are shown as
        short dashed lines, $\psipto\piz X$ backgrounds are in dotted lines,
        continuum in long dashed lines, and $\psipto (\gam_{FSR})X$ in
        short dash-dot-dotted lines.}
    \end{center}
\end{figure}

The signal yields are extracted from an unbinned maximum
likelihood fit to the $M_{VV}^{3C}$ distribution. The signal shape
is obtained from MC simulation, following $BW(m_{0},\Gamma)\times
E_{\gam}^{3} \times damping$, where $m_{0}$ and $\Gamma$ are the
mass and width of the Breit-Wigner for signal and $\chic{J}$,
$E_{\gam}^{3}$ is the cube of the radiative photon energy, which
is necessary in an $E1/M1$ radiative transition, and $damping$
stands for a damping function used to damp the diverging tail
caused by the $E_{\gam}^{3}$ at lower mass region (corresponding
to a higher energy radiative photon). One damping function used by
KEDR~\cite{damp_KEDR} is defined as \(
\frac{E_{0}^{2}}{E_{\gam}E_{0}+(E_{\gam}-E_{0})^{2}}\), where
$E_{0}$ is the most probable energy of the transition photon. It
is also necessary to convolute this with a Gaussian function
$G(\mu,\sigma)$ to take the mass resolution difference between MC
simulation and data into account. The mean ($\mu$) and standard
deviation ($\sigma$) are free parameters for the $\chic{J}$
signals.  For $\etacp$, they are fixed to the values extrapolated
from $\chic{J}$ with a linear assumption.
In the fit, the estimated backgrounds from $\psipto\piz X$ and the
continuum are fixed. The shape of the $\psipto(\gam_{FSR})X$
background comes from the MC simulation. The fraction of MC data
with an $FSR$ photon is scaled by the factor $f_{FSR}$ to estimate
the fraction of data with $FSR$ background. Figure~\ref{fit_VV} shows
the final fitting results to the 3C mass spectrum.
 The values of $\chi^{2}/ndf$ are 0.68 and
0.72 for $\rho^{0}\rho^{0}$ and $K^{*0}\bar{K}^{*0}$,
respectively, indicating good fits. The numbers of $\etacp$ events
 obtained are $6.5\pm6.4$ and $6.9\pm4.8$ for $V$ = $\rho^{0}$
 and $K^{*0}$, respectively. No fit is performed for
$\phi\phi$, since there is only one $\etacpto \phi\phi$ candidate
event in the signal region.

The systematic uncertainties related to tracking, photon
reconstruction, PID and the kinematic fit are estimated with
specially selected control samples~\cite{sys_bes3}.
An efficiency can be defined as the ratio of $\chic{J}$ yield for $VV$
with the $V$ mass requirement to that without this requirement.
The exact same method is applied to MC and
the difference in the efficiency between MC simulation and data is
taken as the corresponding systematic uncertainty caused by the
$V$ mass requirement, with the statistical error included.  An
alternative damping function was used by CLEO~\cite{damp_CLEO}, \(
{\rm exp}(-E_{\gam}^{2}/(8\beta^{2})) \), which is inspired by the
overlap of wave functions, with $\beta=65.0\pm 2.5\MeV$ from
fitting the $\jpsito\gam \eta_c$ photon spectrum. The difference
caused by the two damping functions is taken as a systematic
uncertainty.  The main backgrounds that may affect our fit result
in the $\etacp$ mass region are the contributions from $FSR$ in
$\psipto\gam_{FSR}X$ and from the continuum. Therefore, the
systematic uncertainty from the background shape is estimated by
changing the $FSR$ and continuum contributions by $1\sigma$. There
are also systematic uncertainties related to the mass and width of
the $\etacp$, which are estimated by comparing the $\etacp$ yields
with the mass and width fixed to the center values or randomly
selected values according to a Gaussian distribution.
Table~\ref{sys_sum} shows a summary of all the systematic
uncertainties.

\begin{table}[htbp]
    \caption{The systematic uncertainties in the measured product
    branching fraction $\mathcal{B}(\psipto\gamma\etacp)\times
    \mathcal{B}(\etacpto VV)$.}\label{sys_sum}
    \centering
    \begin{tabular}{cccc}
        \hline\hline
        Source  & $\rho^{0}$ & $K^{*0}$ & $\phi$ \\
        \hline
        Background                  (\%)  & 14.9& 9.9 & 0.0 \\ 
        Tracking                    (\%)  & 8.0 & 8.0 & 8.0 \\ 
        Photon reconstruction       (\%)  & 1.0 & 1.0 & 1.0 \\ 
        Particle ID                 (\%)  & 8.0 & 8.0 & 8.0 \\ 
        4C fit ($\chi^2$ selection)       (\%)  & 4.0 & 4.0 & 4.0 \\ 
        $V$ mass selection requirement    (\%)  & 2.6 & 1.1 & 1.6 \\ 
        Damping function            (\%)  & 40.5& 10.0& 0.0 \\
        Mass and width of $\etacp$  (\%)  & 6.6 & 5.8 & 0.0 \\
        Number of $\psip$           (\%)  & 4.0 & 4.0 & 4.0 \\ 
        \hline
        Total                       (\%)  & 45.6 & 19.9 & 12.8 \\
        \hline\hline
    \end{tabular}
\end{table}

As there is no significant $\etacp$ signal in any of the three
final states, we determine upper limits on the
$\psipto\gam\etacpto \gam VV$ production rates. We assume all the
signal events from the fit are due to $\etacpto VV$, neglecting
possible interference between the signal and nonresonant
contributions. The probability density function (PDF) for the
expected number of signal events is smeared with the systematic
uncertainties (by convolution). For $V=\rho^{0}$ and $K^{*0}$, the PDF is taken to be
the likelihood distribution in fitting the invariant mass
distributions in Fig.~\ref{fit_VV} by setting the number of
$\etacp$ signal events from zero up to a very large number. For
$V=\phi$, the one event in the $\etacp$ mass region is taken as
signal for simplicity, and the PDF is assumed to be a Poisson
distribution.

The upper limit on the number of events at the 90\% C.L., $N^{up}_{\gam VV}$, corresponds to
${\int_0^{N^{up}_{\gam VV}}{\rm PDF}(x)dx}/{\int_0^{\infty}{\rm
PDF}(x)dx}=0.90$ on the smeared PDF. The left half of
Table~\ref{Br_total} shows $N^{up}$, the efficiencies from MC
simulation, and the upper limits on the product branching fraction
$\mathcal{B}$($\psipto \gam \etacp)\times\mathcal{B}(\etacpto
VV)$. Using $\mathcal{B}(\psipto \gam\etacp) = (4.7\pm 0.9\pm
3.0)\times 10^{-4}$~\cite{etacp_bes3}, the corresponding upper
limits on $\mathcal{B}(\etacpto VV)$ are listed in the right half
of Table~\ref{Br_total}. In calculating $\mathcal{B}^{up}(\etacpto
VV)$, the error on $\mathcal{B}(\psipto \gam\etacp)$ is taken as a
systematic uncertainty to smear the PDF. The theoretical
predictions~\cite{theo_VV} on branching fractions for $\etacpto
VV$, which are calculated with $\Gamma_{\etacp}=10.4\pm
4.2\MeV$~\cite{width}, are also listed in Table~\ref{Br_total}.

\begin{table*}[hbt]
    \caption{\label{Br_total}From left to right, they are efficiency, upper limits at the 90\% C.L. on the yield,
    product branching fraction $\mathcal{B}(\psipto
    \gam\etacp)\times\mathcal{B}(\etacpto VV)$, $\etacp$ decay
    branching fraction $\mathcal{B}(\etacpto VV)$, and theoretical
    predictions from Ref.~\cite{theo_VV}.}
    \begin{center}
        \begin{tabular}{lccc|cc}
            \hline\hline
            $V$ & $\varepsilon$ (\%) & $N_{\gam VV}^{up}$ &
            $\mathcal{B}^{up}(\psipto\gam\etacpto\gam VV)$ ($10^{-7}$)&
            $\mathcal{B}^{up}(\etacpto VV)$ ($10^{-3}$) & $\mathcal{B}^{theory}(\etacpto VV)$ ($10^{-3}$)\\
            \hline
            $\rho^{0}$ & $14.3$ & 19.2
            & 12.7 & 3.1  & 6.4 to 28.9 \\
            $K^{*0}$ & $16.5$ & 15.2
            & 19.6 & 5.4  & 7.9 to 35.8 \\
            $\phi$ & $19.9$ & 3.9
            & 7.8 & 2.0 & 2.1 to 9.8 \\
            \hline\hline
        \end{tabular}
    \end{center}
\end{table*}

In conclusion, no obvious $\etacp$ signal was observed in decays
into vector meson pairs: $\rho^{0}\rho^{0}$, $K^{*0}\bar{K}^{*0}$,
and $\phi\phi$. The upper limits on the product branching fraction
$\mathcal{B}(\psipto \gam \etacp)\times\mathcal{B}(\etacpto VV)$
and $\etacp$ decay branching fraction $\mathcal{B}(\etacpto VV)$
are determined. These upper limits are smaller than the lower
bounds of the theoretical predictions~\cite{theo_VV}, although the
difference is very small for $\etacp \to \phi \phi$.

The BESIII Collaboration thanks the staff of BEPCII and the
computing center for their hard efforts. This work is supported in
part by the Ministry of Science and Technology of China under
Contract No. 2009CB825200; National Natural Science Foundation of
China (NSFC) under Contracts No. 10625524, No. 10821063, No. 10825524,
No. 10835001, No. 10935007; the Chinese Academy of Sciences (CAS)
Large-Scale Scientific Facility Program; CAS under Contracts No.
KJCX2-YW-N29, No. KJCX2-YW-N45; 100 Talents Program of CAS; Istituto
Nazionale di Fisica Nucleare, Italy; Siberian Branch of Russian
Academy of Science, joint project No. 32 with CAS; U. S.
Department of Energy under Contracts No. DE-FG02-04ER41291,
No. DE-FG02-91ER40682, No. DE-FG02-94ER40823; University of Groningen
(RuG) and the Helmholtzzentrum fuer Schwerionenforschung GmbH
(GSI), Darmstadt; WCU Program of National Research Foundation of
Korea under Contract No. R32-2008-000-10155-0.


\begin{thebibliography}{**}
\bibitem{svv_model}The amplitudes and phases for H+, H0, and H- are 1.0, 0.0, 0.0, 0.0, -1.0, 0.0, which are calculated by requring of $P$-parity conservation for the allowed components of helicity amplitudes.
\bibitem{rec_mass}The $\pipi$ recoil mass is defined as $M_{\pipi}^{\rm recoil}=\sqrt{(E_{\psip}-E_{\pi^{+}}-E_{\pi^{-}})^2-
    (\overrightarrow{p}_{\psip}-\overrightarrow{p}_{\pi^{+}}-\overrightarrow{p}_{\pi^{-}})^2}$, where $E$ and $\overrightarrow{p}$ are
    energy and momentum for a particle.
\bibitem{f_Novo}The Novosibirsk function is defined as $f(m_{ES}) = A_S \exp ( -0.5 { \ln^2 [1 + \Lambda \tau \cdot (m_{ES} - m_0) ] / \tau^2 + \tau^2 } )$, where $\Lambda = \sinh ( \tau \sqrt{\ln 4} ) / ( \sigma \tau \sqrt{\ln 4} )$, the peak position is $m_0$, the width is $\sigma$, and $\tau$ is the tail parameter.
\end{thebibliography}
\end{document}